\documentclass{nature}

\usepackage{graphicx}
\usepackage[dvipsnames]{xcolor}
\usepackage{changes}
\usepackage[left=2.0cm, right=2.0cm, top=2.00cm, bottom=2.00cm]{geometry}
\usepackage{tikz}
\usetikzlibrary{shapes}
\usepackage{graphicx}				
\usepackage{dcolumn}		
\usepackage{bm}				
\usepackage{float}
\usepackage{amsmath}
\usepackage{amssymb}
\usepackage{amsfonts}
\usepackage{times}
\usepackage{tabularx}
\usepackage{xcolor,colortbl}
\usepackage{xcolor}
\usepackage[11pt]{moresize}
\usepackage{gensymb}

\usepackage{graphicx}
\usepackage[dvipsnames]{xcolor}
\usepackage{tikz}
\usetikzlibrary{shapes}

\newcommand{\markerone}{\raisebox{0.5pt}{\tikz{\node[draw,scale=0.65,circle,fill=black!20!ProcessBlue](){};}}}
\newcommand{\markertwo}{\raisebox{0.5pt}{\tikz{\node[draw,scale=0.65,circle,fill=black!20!BurntOrange](){};}}}
\newcommand{\markerthree}{\raisebox{0.5pt}{\tikz{\node[draw,scale=0.65,circle,fill=black!20!ForestGreen](){};}}}

\makeatletter
\let\saved@includegraphics\includegraphics
\AtBeginDocument{\let\includegraphics\saved@includegraphics}
\renewenvironment*{figure}{\@float{figure}}{\end@float}
\makeatother

\definecolor{commentColor}{rgb}{0.0,0.6,0.0}
\definecolor{todoColor}{rgb}{0.9,0.0,0.0}

\title{Walker-like Domain Wall breakdown in layered Antiferromagnets driven by staggered spin-orbit fields}

\author{Rub\'{e}n M. Otxoa$^{1,2,*}$, P. E. Roy$^{1}$, R. Rama-Eiroa$^{2,3}$, J. Godinho$^{5}$, K. Y. Guslienko$^{3,4}$ \& J. Wunderlich$^{1,5}$}

\begin{document}

\maketitle

\begin{affiliations}
 \item Hitachi Cambridge Laboratory, J. J. Thomson Avenue, CB3 OHE, Cambridge, United Kingdom
 \item Donostia International Physics Center, Paseo Manuel de Lardizabal 4, Donostia-San Sebastian 20018, Spain
 \item Depto. F\'{i}sica de Materiales, Universidad del Pa\'{i}s Vasco, UPV/EHU, 20018 San Sebasti\'{a}n, Spain
 \item Ikerbasque, The Basque Foundation for Science, 48013 Bilbao, Spain
 \item Institute of Physics ASCR, v.v.i., Cukrovarnicka 10, 162 53 Praha 6, Czech Republic
 \end{affiliations}

\begin{abstract}

Within linear continuum theory, no magnetic texture can propagate faster than the maximum group velocity of its spin waves. Here we report a transient regime due to the appearance of additional antiferromagnetic textures that breaks the Lorentz translational invariance of the magnetic system by atomistic spin dynamics simulations. This dynamical regime is akin to domain wall Walker-breakdown in ferromagnets and involves the nucleation of an antiferromagnetic domain wall pair. Subsequently, one of the nucleated 180$^{\circ}$ domain wall creates with the original domain wall a 360$^{\circ}$ spin-rotation which remains static even under the action of the spin-orbit field. The other 180$^{\circ}$ domain wall becomes accelerated to super-magnonic speeds. Under large spin-orbit fields, multiple domain wall generation and recombination is obtained which may explain the recently experimentally observed current pulse induce shattering of large domain structures into small fragmented domains and the subsequent slow recreation of large-scale domain formation prior current pulse.

\end{abstract}

In a general physical context, a topological defect is characterised by a local/core region where order is highly altered whereas far from that core region the order parameter varies smoothly in the space\cite{chaikin2000principles}. Topological defects can be found in a variety of fields such as in superfluid helium\cite{anderson1973anisotropic} named vortices, in periodic crystal structures, they are called dislocations\cite{levine1985elasticity,socolar1986phonons,kosterlitz1972long} and in magnetism, magnetic domain walls\cite{bloch,neel}, vortices\cite{feldtkeller,bogdanov2001chiral} and skyrmions \cite{rossler2006spontaneous}. In elementary dislocations, the underlying reason why these defects are called topological is because they cannot be made to disappear by a continuous deformation of the order parameter in infinite extended films. The measure of the distortion is associated to the so-called Burgers vector whose magnitude expresses the strength of the dislocation. The rule consists in taking a closed path within a chosen portion of the crystal lattice, moving in steps of the lattice parameter and keeping the number of steps in each direction equal as the circuit is completed. If the section of crystal is defect-free, the end point will then coincide with the starting point. If however there is a defect contained within the circuit, then the end-point will not be equal to the starting point. The vector from the end-point to the starting-point is then the Burgers vector and its magnitude is used as a measure of the strength of the dislocation. The Burgers vector is quantised to integer multiples of the atomic spacing of the lattice vectors\cite{weertman1965elementary}. Continuum-based textures such as magnetic textures like domain walls, vortices and skyrmions are often characterised by the so called winding number whose value stands in direct relation to the magnitude of the normalised Burgers vector. However, topology is not at odds with instability. A relevant example of the previous is the plastic deformation and its role in dislocation dynamics\cite{nadgornyi1988dislocation}. Under low applied stresses, there is a quasilinear dependence between the dislocation velocity and the applied stress described by the linear elasticity theory\cite{hirth1982theory}. However, dislocation mobility is not restricted to subsonic regime under high strain-rates. Surpass the sonic barrier was first postulated theoretically\cite{eshelby1949uniformly,stroh1962steady,frank1949one} and then confirmed by atomistic simulations in bcc tungsten\cite{gumbsch1999dislocations,li2002dislocation} and in iron screw dislocations\cite{marian2004dynamic} by means of introducing secondary kinks via mother-daughter nucleation process with opposite Burgers vector\cite{vandersall2004supersonic}.

Within the micro magnetic approximation, in a two-dimensional infinite magnetic system, the order parameter that breaks the symmetry is reflected by an average quantity: the magnetisation $\vec{m}(x,y)$ at each site in a given lattice. The connection between topology and magnetic media and its role in dynamics became apparent by the pioneering work of several authors\cite{thiele1973steady,malozemoff2016magnetic,odell1981ferromagnetodynamics} when studying the dynamics of magnetic topological defects. The analytical approach consists of assuming the centre of the texture as a dynamical variable where the equation of motion can be derived using Lagrangian formalism. Concretely, in ferromagnets, domain walls (DW) are defined by the transition region that separates two homogeneously and opposite magnetised regions. In a similar fashion as for dislocations, a quantised topological invariant is defined for the magnetisation field\cite{belavin1975pis}. By framing the magnetic texture on a closed path, it allows to define a map of the magnetisation configuration. Distinct topological textures are characterised by different winding number, $w$, which counts the number of times the magnetisation is wrapped onto itself\cite{Braun:2012kw}. In the case of Bloch walls separating magnetic domains in ferromagnets topological defects arise in the form of N\'{e}el lines\cite{thiaville1990neel}. Since the nineties, the study of DW mobility in ferromagnets has been the flagship in Spintronics because of its potential in data storage\cite{parkin2008magnetic,tatara2008microscopic} and logic devices\cite{allwood2005magnetic}. However, a major difficulty  to implement such technological proposals arise as a result of the intrinsic instability the DW structure presents when it overcomes a certain threshold velocity. In this Article we show that this interesting feature of domain walls is present irrespective of whether the magnetic material is a ferromagnet or antiferromagnet. In ferromagnets, this regime is known as the Walker limit\cite{schryer1974motion} and it is characterised by a DW mobility that is the result of the combination of translational and oscillatory dynamics\cite{mougin2007domain} leading to to periodic alteration of the DW chirality. Unfortunately, the Walker breakdown (WB) occurs at a low domain wall velocities and imposes a limit on a reliable control of the DW motion. An alternative to prevent the WB consists of using nanotubes or nano cylinders\cite{hertel2004magnetization,yan2011fast}. These particular geometries have unveiled the spin analog of the Cherenkov effect which involves the emission of electromagnetic waves by charge particles moving at speeds faster than the speed of light in the given medium\cite{vcerenkov1937visible}. In the case of antiferromagnets, several studies have provided significant insights on DW motion in antiferromagnets where high speeds have been predicted thanks to prevention of the WB\cite{shiino2016antiferromagnetic,gomonay2016high} yet there is an upper theoretical limit which can not be surpassed. Interestingly, the manner in which the spin-space is conformed in antiferromagnets obeys the relativistic kinematics of special relativity where the analogue role played by the photons in conventional space is played by the magnons in a magnetic system. Consequently, by comparison, the ultimate limiting velocity of any texture is given by the maximum group velocity of the magnons, $v_{\text{g}}$. An interesting question involves whether, under high-strain, in the same fashion as it occurs for dislocations under plastic deformation, super-magnonic regime of motion could occur through a sequence of additional nucleation textures mimicking the so-called WB. This behaviour would be akin to supersonic motion of dislocations through nano-twin formation. To elucidate the existence of such type of dynamics, we combined one-dimensional atomistic spin dynamics simulations of a the layered antiferromagnet, Mn$_{2}$Au\cite{roy2016robust}, with one-dimensional analytical theory. In order to induce magnetisation dynamics in Mn$_{2}$Au (see \textbf{Fig. 1 a}), we make use of the predicted staggered field-like torque in such crystal structures \cite{vzelezny2014relativistic}, where the effective magnetic field resulting from a staggered induced spin-density $\vec{H}^{\text{so}}$, possesses opposite signs at each sub-lattice and gives rise a spin-orbit torque (see $\textbf{Fig. 1 b}$). Some representative dynamics of the temporal evolution of DW position is obtained by extracting the spatiotemporal evolution of the winding number density shown in \textbf{Fig. 1 c, d}. These spatiotemporal evolutions were obtained by firstly ramping the Spin-Orbit field (SO-field) from 0 to its maximum in a time interval of 10 ps and then it is kept constant till the end of the simulation. The intrinsic inertial character of the translational motion of the DW is revealed by a non-linear trend that appears within the first 4-5 picoseconds. The ballistic nature of the DW motion is revealed by the narrowing of the DW as it approaches $v_{\text{g}}$ according to the relativistic kinematics framework.  The DW width at a given velocity is given by: $\Delta(v)=(1-v^{2}/v^{2}_{\text{g}})^{1/2}\Delta_{0}$, where $\Delta_{0}$=19.8 nm is the DW width at rest. This contraction translates into an increase of the DW exchange energy.  It is noteworthy to comment that in conventional relativity theory, the contraction observed from non-inertial reference frame does not result in any incremental (or detrimental) interaction among the constituents of the object. First, the DW is accelerated to a speed that approaches to $v_{\text{g}}$. After this transient dynamics, two distinct behaviours are obtained depending on the applied SO-field. For low SO-fields (see \textbf{Fig. 1 c}), a saturation effect of the DW speed is observed, characterised by the slope of the maximum winding number density position (DW position), $q$, with respect to time, $t$ with  $v=\partial_t q$.  A linear variation of the DW speed as a function of the SO-field is consistent with previous reports \cite{gomonay2016high} (see Supplementary Information). In such a case, the energy dissipation induced by the viscous damping inhibits the DW to move faster as reported in previous work\cite{shiino2016antiferromagnetic}. In contrast, at an onset of 65 mT, a different dynamical behaviour is observed which can be defined as the critical stress at which the DW suffers a threshold deformation which leads to the breeding of a DW-pair with trivial composite winding number, $w$, preserving the overall winding number (see $\textbf{Fig. 1 d, bottom panel}$) . This strongly nonlinear regime which involves the generation of additional particles has been reported under several conditions in ferromagnets, for instance, domain wall WB mediated by the vortex-antivortex generation\cite{yan2011fast}, vortex-core reversal in spin-torque oscillators in nanodots\cite{guslienko2008dynamic} and nanocontacts \cite{petit2012commensurability}. However, its occurrence in antiferromagnets has never been reported nor proposed. In ferromagnets, the physical origin for the appearance of new magnetic textures, concretely in the case of vortex-core reversal has been associated to the emergence of a magnetic field whose origin is purely dynamical and can be derived from the kinetic part of the Lagrangian density, $\mathcal{L}=\mathcal{L}_\text{kin}-V$. In our system, the DWs are located in the ferromagnetic basal atomic planes where the coupling between the adjacent planes is antiferromagnetic. We can derive the temporal evolution of the kinetic field using the Lagrangian formalism.

\begin{equation}
\vec{h}_\text{kin}(t)=\frac{1}{M_\text{s}}\frac{\partial \mathcal{L}_\text{kin}}{\vec{m}(t)},
\end{equation}

where $M_\text{s}$ is the sub-lattice magnetisation. Although this emergent field has already been proposed to explain the reversal of the vortex core polarity in ferromagnets where the dynamics are gyrotropic, we use this formalism also to translational motion of DWs in layered antiferromagnets. In antiferromagnets due to the relativistic nature of the dynamics, the DW already suffers a substantial deformation due to its translational velocity. Interestingly, the extension and the kinetic field profile central location with respect to the moving DW depends upon the dynamical regime the DW is at a given instant (see \textbf{Fig. 2 a-d}). For instance, during the first 4-5 ps of ultra-fast acceleration, the kinetic field is located behind the DW and smoothly transitions from the back to the center and then to the forefront of the DW while mimicking the DW profile. The maximum component of the kinetic field is found to be along the $\hat{z}$-direction and can reach values up $\sim$35 T in absolute value. Contrary to ferromagnets where the magnetic texture absorbs energy from the kinetic field, we attribute the nucleation of the DW pairs to the torque exerted by the kinetic field onto the local magnetisation promoting a reverse magnetic domain along the $-\hat{x}$ with an approximate extension just before the DW-pair nucleation of the exchange length, $l_\text{ex}$=(A/K)$^{1/2}$, where $A$ is the effective exchange stiffness per unit area and $K$ is the uniaxial anisotropy (see Table 1). The evolution of this reverse magnetic domain is govern by the competition between the exchange and anisotropy energies which results into breaking the reverse magnetic domain into a DW-pair with each of its constituents holding an opposite winding number (thus the sum of their windings is zero). We identify the time involved in the nucleation to be around 1 ps. Different type of magnetic distribution could in principle appear obeying the boundary conditions over the generated magnetic domain, however, we observed that the initially moving DW and the nucleated DW located just in front share the same winding number. The origin of this arrangement can be attributed to the minimisation of the Zeeman energy now comprised of the SO-field and the kinetic field (see Supplementary material). Therefore, while the torque provided by the kinetic field leads to a reverse magnetic domain in front of the moving DW with the appropriate extension to accommodate two DWs, the competition between the magnetic energies and the minimisation of the Zeeman energy gives rise to the generation of a DW-pair with a specific ordering.

The temporal evolution of the system after the DW-pair is present in the system is dramatically affected. Hereafter, we call the initially moving DW, DW$_1$ and the nucleated DWs, DW$_2$ and DW$_3$ being DW$_2$ always the closest to DW$_1$. As can be seen in \textbf{Fig. 3 a}, immediately after the nucleation, DW$_1$ and DW$_2$ perform few oscillations and then get stuck evolving in time but not in space. As DW$_1$ and DW$_2$ have the same winding number, they repel each other due to the ferromagnetic exchange interaction (see Fig. \textbf{Fig. 3 a}). However, the region in between the pair (DW$_1$ and DW$_2$) has opposite orientation with respect to $\vec{H}_{\text{so}}$ and therefore, in order to reduce the Zeeman energy provided by the SO-field (Note that the kinetic field is zero as the DWs are static), the distance between DW$_1$ and DW$_2$ has to be minimised. Consequently, as a result of the competition between the two forces, a local minimum appears that corresponds to a certain distance between the DW-pair. We verified this interpretation by calculating the energy barrier related to the distance among two DW$_1$ and DW$_2$ (see \textbf{Fig. 3 b}). The energy minimum corresponds to a given separation between the DWs' center, which depends on the magnitude of the SO-field (see Supplementary material). From numerical simulations, we obtain a stable distance of 32 nm whereas from analytical calculation the distance is 29 nm for $\vec{H}_{\text{so}}$= 65 mT.

Another observation is that DW$_3$ propagates forward by undergoing a velocity boost that surpass the maximum group velocity of the magnons. This regime is transient and last for several tens of picoseconds which is sufficient for DW$_3$ to cover over three microns. This mobility regime can be explained as follows. While the DW-pair is nucleated, DW$_1$ and DW$_2$ repel each other thanks to the ferromagnetic exchange interaction between them leading to a potential motion in opposite directions for each DW. Despite this fact, they barely move (See inset \textbf{Fig. 3 a}) and therefore, all this latent momentum is transferred from DW$_1$to DW$_3$\cite{tveten2014antiferromagnetic}. Hence, the DW$_3$ mobility at the moment of the nucleation consist of the velocity provided by the DW$_1$, in addition to the boost coming from the momentum transfer due to the repulsion between DW$_1$ and DW$_2$. Using collective coordinates approach combined with the Rayleigh dissipation function and the Euler-Lagrange equation of motion, we obtain the DW$_3$ velocity as a function of its distance with respecto the de DW-pair to be

\begin{equation}
v_{3}=\frac{\gamma}{\alpha}H_{\text{so}}\Delta_1+\frac{2 \gamma A}{\alpha M_{\text{s}}\Delta^{2}_{\text{DW}_{3}}}\left[x(t)\text{coth}\left(\frac{x(t)}{\Delta_{\text{DW}_{3}}}-\Delta_{\text{DW}_{3}}\right)\right]\text{csch}\left(\frac{x(t)}{\Delta_{\text{DW}_{3}}}\right),
\end{equation}

\noindent where $\alpha$ is the Gilbert damping, $\Delta_{\text{DW}_{1}}$ and $\Delta_{\text{DW}_{3}}$ are the DWs' width for DW$_1$ and DW$_3$ at the nucleation extracted from the atomistic simulations respectively and $x(t)$ is the distance between DW$_1$ and DW$_3$. The maximum speed according to Eq. 2 is circa 133 km/s which corresponds to a distance of 16.1 nm between the DW$_1$ and DW$_3$. From numerical simulations, the extracted distance between DW$_1$ and DW$_3$ is 17 nm when the nucleation starts occurring and the extracted speed value for DW$_3$ is around 177 km/s which is circa 4 times the maximum group velocity that magnons can theoretically attain in Mn$_2$Au (see Supplementary Material). It is appealing to think that the special theory of relativity is being violated as no magnetic texture is by principle allowed to propagate faster than the maximum velocity of the magnons in the spin space. However, as soon as the DW-pair (DW$_2$ and DW$_3$) is present in the system additional interactions need to be taken into account which result into a Lagrangian that locally is not Lorentz invariant and therefore, the constraint on the maximum speed must be lifted up. Besides, when DW$_3$ enters into a super-magnonic regime of motion, there is an explosion of spin-waves propagating together but never surpassing DW$_3$ (see \textbf{Fig. 2 d and Supplementary Material}). Analogous effect has been reported in mechanical systems such as in edge dislocation dynamics under shear stress where spontaneous emission of radiation has also been observed\cite{peng2019supersonic,nosenko2007supersonic}. We attribute the origin of the emitted spin-waves to a mixture of the so-called Bremsstrahlung effect\cite{tesla} (or breaking radiation) and the spin-Cherenkov effect\cite{yan2013spin}. The Bremsstrahlung effect arises due to the deceleration of a charge particle. Therefore, one could speculate that such staggering deceleration could lead to excitations of the spin medium. As shown in \textbf{Fig. 3 d}, DW$_3$ not only surpass the maximum group velocity of the magnons but it also enters in a mobility regime where it moves faster than the phase velocity of the magnons. It is difficult in our case to clearly separate the existence of Bremsstrahlung and spin Cherenkov radiation. We note in Figures \textbf{1 d} and \textbf{3 a} that the number of oscillations of DW$_1$ and DW$_2$ equals the number of spin-wave ripples travelling with DW$_{3}$ and further, that the decay of these ripples appear to stretch over a rather long time. Therefore, one possibility for co-existence of both types of radiation could be that the Spin-Cherenkov-originated spin-waves acts in an anti-damping fashion onto the Bremsstrahlung in conjunction with the observation that the radiative ripples to not propagate in opposite direction to DW$_{3}$. 
It is pertinent to explore how the system reacts when the applied SO-field as the system will have at its disposal additional pumped energy. We start by injecting an electric current with the time profile illustrated in \textbf{Fig. 4 a}; The electric current has a rising time of 5 ps up to a peak value of $H_{\text{so}}(t)=100$ mT. The value of $H_{\text{so}}$ is kept constant for the following 50 ps before reducing it to zero, with a falling time of 5 ps. The SO-field is set to zero for 50 ps before starting this pattern again three times in succession. The coloured circle \markerone $\,$ in \textbf{Fig. 4 a}, represents the maximum $\pi/2$-windings, i/e/ the number of DWs in each pulse. In each cycle, we can observe an avalanche of DW-pairs. Once the primal DW nucleates the first DW-pair, the DW that propagates from the DW-pair (DW$_1$ and DW$_2$) at supermagnonic speeds becomes a new breeder and gives rise to new DW-pair. This phenomenon repeats for a 13 times leading to the appearance of 26 additional DWs in the system preserving at all times the overall topological charge. It is worth noticing that the DWs once nucleated do not reorganised into a more stable configuration while the SO-field is maximum giving rise to DW-lattice-like structure whose inter-DW distance depends upon the SO-field pulse pattern. The average DW density per pulse is 0.013 DW/nm and takes approximately 20 ps to generate the whole DW lattice. \markertwo $\,$ denote the winding density number once the SO-field starts decreasing. Once this occur, the lattice decompresses which, in combination to the attractive exchange interaction among DWs with opposite winding number, leads to recombination of some DWs. We observed a similar value for \markertwo $\,$ with pulse number meaning that the same number of DWs recombine once the SO-reaches 0 mT. The winding density number represented by \markerthree $\,$ is extracted once the SO-field is about to rise up again. It can be seen from \textbf{Fig. 4a} that \markerthree $\,$ increases in each pulse. This implies that the accumulated number of remnant DWs in the system after each pulse increases meaning that not all the generated DWs annihilate. 
After the 4th pulse, the number of DWs in the system is 8 with a spacing among the DWs that ranges from hundreds of nanometers to few micrometers. Initially the system contained  two magnetic domains (separate by a single DW) and after 4 pulses the number of magnetic domains,$N_{\text{m}}$, has increased by $N_{\text{m}}=N_{\text{dws}}+1$, where ,$N_{\text{dws}}$, accounts for the number of DWs in the system. More complex DW-crystal lattices can be obtained by varying the pulse duration and the sign of the SO-field (see Supplementary material) showing the rich dynamics in these type of systems. We wish at this point to also stress an observations (not shown here), that there appears to be no clear correlation between the strength of $H_{\text{so}}$ and the number of DWs nucleated as well the point in time and space where they are nucleated. This opens up the possibility that this system under the current excitation circumstances could potentially exhibit chaotic behaviour. This will be the topic of future work however. We note at this stage, that recent work on current-induced resistance changes in the antiferromagnet CuMnAs was, in conjunction with imaging attributed to the fragmentation and recovery of the domain structure\cite{Wornle,kaspar}. In particular they observed a gradual increase in the resistance with repeated pulsing and a slow relaxation of the resistance towards lower values than after the last pulse when turning off the excitation completely. 
However, provided that finite size-effects can be excluded and that such a system observes global conservation of the total winding number, it should be possible to return to the original state by applying a reverse field below the AFM Walker-breakdown field, in order to recover completely the initial state of resistance, in other words, a complete reset of the system.
Further, it is noteworthy that the relaxed resistance value after a long waiting time was higher than the original starting resistance. This could mean that residual domain walls are present. If the interpretation that the observed resistance changes is due to domain fragmentation and recovery/recombination is correct we cannot help but to speculate about possible measurements of our system under study. Although the mechanism of domain fragmentation in our case is due to the AFM Walker-breakdown  in a perfect crystal and the results in the reported experimental work likely has heat as the main cause of fragmentation this means that that such resistance variations in analogous systems where pinning and heating effects can be ruled out could be used as an indirect detection method for the occurrence of the AFM Walker-breakdown process and its associated generation of supermagnonic textures.  
It must be pointed out that for this particular pulse ramp-time no signature of DW-pair nucleation is obtained well after the SO-field has reached $H_\text{crit.}$= 65 mT, meaning that the critical field depends upon the ramping time conditions rather than the absolute value of the energy pumped into the system. \textbf{Fig. 4 b} shows the transition from a system with a DW-lattice with 13 DW-pairs towards a system with only one DW-pair due to the multiple recombination when the SO-field is zero. It is noted that there is a distribution of distances that separate consecutive DWs with opposite winding number. This gives rise to a manifold of recombination times as shown in \textbf{Fig. 4 c} . Analytically, it is possible to obtain the recombination time as a function of the distance by assuming that the decompression introduced by the absence of SO-field results in a transition phase from a DW-lattice to a DW-gas for a transient time. Therefore, the recombination time of a given DW-pair is only governed by intrinsic parameters such as, the exchange interaction, the damping and the distance between the DWs ruling out the effect of the presence of the rest of DW-pairs in the system. We can derive the recombination time from a general expression of the exchange interaction between two domain wall pairs (see Supplementary material) with opposite winding number as
 
\begin{equation}
\tau_{\text{r}}(l)=\int_{t_0}^{t_{\text{r}}}dt=-\frac{\alpha M_{\text{s}}\Delta}{2 \gamma A} \int_{l}^{l_0}dx\frac{1}{\left[\Delta-x \text{coth}\left(\frac{x}{\Delta}\right)\right]\text{csch}\left(\frac{x}{\Delta}\right)},
\end{equation} 

\noindent where $\Delta$ is the DW-width of each DW, $l$ represents the distance between the DWs that will recombine at $t_0$, $l_0=$1 nm and $t_{\text{r}}$ represents the recombination time. We note that the predicted recombination time reproduce quantitatively the recombination time values extracted from the atomistic spin dynamic simulations which validates the hypothesis of 1D DW-pairwise gas approximation. For distances in the range of few hundreds of nanometers the recombination time is in the range of few picoseconds as the exchange interaction is a short range interaction. However we can see from \textbf{Fig. 4 b} that there are DWs remaining in the system whose separation is in the range of few microns. For a given distance of 1.8 $\mu$m, the expected recombination time lies in the range of $\sim$ 6 days. This suggests that in an ideal scenario where pinning effects, thermal fluctuation effects are not present, the stability of such a configuration is granted thanks to the absence of long-range interaction. We can observed in \textbf{Fig. 4 b} that once the SO-field is set to zero, each of the recombination processes produces an excitation in the continuum spectrum. This highly nonlinear process yields to a deformation amplitude that oscillates in time with a very precise frequency. The emergence of this breather-like excitation is due to the excess of kinetic energy carried by each DW involved in the collision whose value is not large enough to escape the attractive potential provided by its anti-particle. By mapping the breather into a simple damped harmonic oscillator, we obtained a good quantitative agreement on the breather lifetime. The breather decay occurs in 20-30 ps and it is governed by Gilbert damping. Moreover, the characteristic frequency of the breather is determined to 544 GHz frequency, which lies within the linewidth of the calculated spin-wave band gap at $H_{\text{SO}}$= 100 mT , i.e. the frequency at zero wave vector, $k$=0 at. We are thus compelled to assign the breather frequency to that of the band-gap.

\section*{Conclusion}
In conclusion, we have studied by ASD and theory, the dynamical properties of staggered field-driven DWs. A Walker-breakdown-like process has been identified whereby the DW nucleates an additional DW pair with a total trivial winding. One of these nucleated domain walls forms a $\pi$ DW together with the original one, whereas the other nucleated DW travels away from the breakdown site as speeds far exceeding the maximum group velocity of spin-waves in the medium, thus achieving supermagnonic speeds. At such high speeds we observe a radiative tail travelling together with the supermagnonic texture. According to the computed dispersion relation, for this system, it is only at supermagnonic speeds where the conditions for spin-Cherenkov radiation is met. Thus one contributing source to the observed radiation, we attribute to spin-Cherenkov radiation. At the same time, we would expect a breaking radiation to be present due to a rapid deceleration of DW{$_1$} and DW$_{2}$. We observe associated oscillations in position of the said DWs as they come to a halt, with the number of oscillations equaling that of the number of radiative ripples travelling with DW$_{3}$, opening the possibility of co-existence of the two types of radiation with the spin-Cherenkov radiation acting in an antidamping manner onto the breaking radiation , which could explain the rather slow decay of the radiative ripples. At higher spin-orbit field magnitudes, both multiple DW nucleations and DW recombinations occur, in order to keep conservation of the global winding number. Provided that the separation between nucleated textures is sufficiently large, residual domain walls can remain in the system after the pulse-excitation is turned off. Provided that the origin of the measured resistance in CuMnAs is of magnetic origin, in particular due to magnetic domain fragmentation\cite{Wornle,kaspar}, our results imply that no Joule heating or temperatures approaching the N\'{e}el temperature are required in order to observe such resistance characteristics.

\newpage
\begin{methods}

We perform numerical simulation via atomistic spin dynamics simulations. For this, the full Mn$_{2}$Au crystal structure is taken into consideration. Two formula units as that in Fig. (1a) in the main text is stacked on top of each and then replicated along the $x$-direction 60000 times. The system has open boundaries along $x$ and $z$ while periodic boundary conditions are imposed along $y$. The time evolution of a unit vector spin at site $i$, $\textbf{S}_{i}$, is simulated by solving the Landau-Lifshitz-Gilbert equation:
\begin{equation}
\label{eq:LLG}
\frac{d\textbf{S}_{i}}{dt} = -\gamma\,\textbf{S}_{i}\times\textbf{H}_{i}^{\text{eff}}-
\gamma\alpha\,\textbf{S}_{i}\times\left(\textbf{S}_{i}\times\textbf{H}_{i}^{\text{eff}}\right), 
\end{equation} 
where $\gamma$ is the gyromagnetic ratio of a free electron (2.21$\times\text{10}^{5}$m/As), $\alpha$ is the Gilbert damping set here to 0.001 and $\textbf{H}_{i}^{\text{eff}}$ is the effective field resulting from all of the interaction energies. The energies taken into account are the three exchange interactions (two antiferromagnetic and one ferromagnetic), magneto crystalline energy contributions and the spin-orbit field. The total energy, $E$ considered is:
\begin{eqnarray}
\label{eq:Energy}
E = -2 \sum_{\langle i<j\rangle}{J_{ij}\textbf{S}_{i}\cdot\textbf{S}_{j}}-
K_{2\perp}\sum_{i}{\left(\textbf{S}_{i}\cdot\hat{\textbf{z}}\right)^{2}}- \nonumber
K_{2\parallel}\sum_{i}{\left(\textbf{S}_{i}\cdot\hat{\textbf{y}}\right)^{2}}- \\
\frac{K_{4\perp}}{2}\sum_{i}{\left(\textbf{S}_{i}\cdot\hat{\textbf{z}}\right)^{4}}-
\frac{K_{4\parallel}}{2}\sum_{i}{\left[\left(\textbf{S}_{i}\cdot\hat{\textbf{u}}_{1}\right)^{4}+
\left(\textbf{S}_{i}\cdot\hat{\textbf{u}}_{2}\right)^{4}\right]}-
\mu_{0}\mu_{S}\sum_{i}{\textbf{S}_{i}\cdot\textbf{H}_{i}^{\text{eff}}}.
\end{eqnarray}
The first term on the right-hand side is the exchange energy where $J_{ij}$ is the exchange coefficient along the considered bonds (see supplementary material). The second and third terms are the uniaxial hard and easy anisotropies of strengths $K_{2\perp}$ and $K_{2\parallel}$, respectively, while the fourth and fifth terms collectively describes tetragonal anisotropy. For the in-plane part of the tetragonal anisotropy, $\textbf{u}_{1}$=$\left[110\right]$ and $\textbf{u}_{2}$=$\left[1\bar{1}0\right]$. Finally, $\mu_0$ and $\mu_{\text{s}}$ are the magnetic permeability in vacuum and the magnetic moment, respectively. We have used $\mu_{\text{s}}= 4\mu_{B}$, with $\mu_{B}$ being the Bohr magneton. It is to be noted that the tetragonal anisotropy was included for sake of completeness as it is present in this material but its role in the high speed dynamics is negligible due to the weak magnitude of its anisotropy constants.
The effective field is then subsequently evaluated at each point in time using Eq. (\ref{eq:Energy}) as $\textbf{H}^{\text{eff}}_{i}=\frac{-1}{\mu_{0}\mu_{S}}\frac{\delta E}{\delta \textbf{S}_{i}}$. 
The system of equations, Eq.(\ref{eq:LLG}) are solved by a fifth order Runge-Kutta Method. Spatio-temporal data is analysed on an extracted 1-dimensional line of the computational domain (see supplementary material). Material constants used are summarised in the following table:
\begin{table}
\begin{center}
	\begin{tabular}{||c c c c c c c c||} 
		\hline
		 $J^{*}k_{\text{B}}^{-1}$[K]&$J_{i1}k_{\text{B}}^{-1}$[K] & $J_{i2}k_{\text{B}}^{-1}$[K] & $J_{i3}k_{\text{B}}^{-1}$[K] & $K_{2\perp}$[J] & $K_{2\parallel}$[J] & $K_{4\perp}$[J] & $K_{4\parallel}$[J] \\ [0.5ex] 
		\hline\hline
		156 & -396 & -532 & 115 & -1.303$\times\text{10}^{\text{-22}}$ & 7$K_{4\parallel}$ & 2$K_{4\parallel}$ & 1.855$\times\text{10}^{\text{-25}}$ \\ 
		\hline
	\end{tabular}
\end{center}

\caption{Literature values for material parameters relevant for modelling the spin dynamics\cite{}. $k_{\text{B}}$ is Boltzmann's constant.}
\label{table:1}
\end{table}
\end{methods}

\newpage


\begin{figure}
\centering
\includegraphics[scale=0.55]{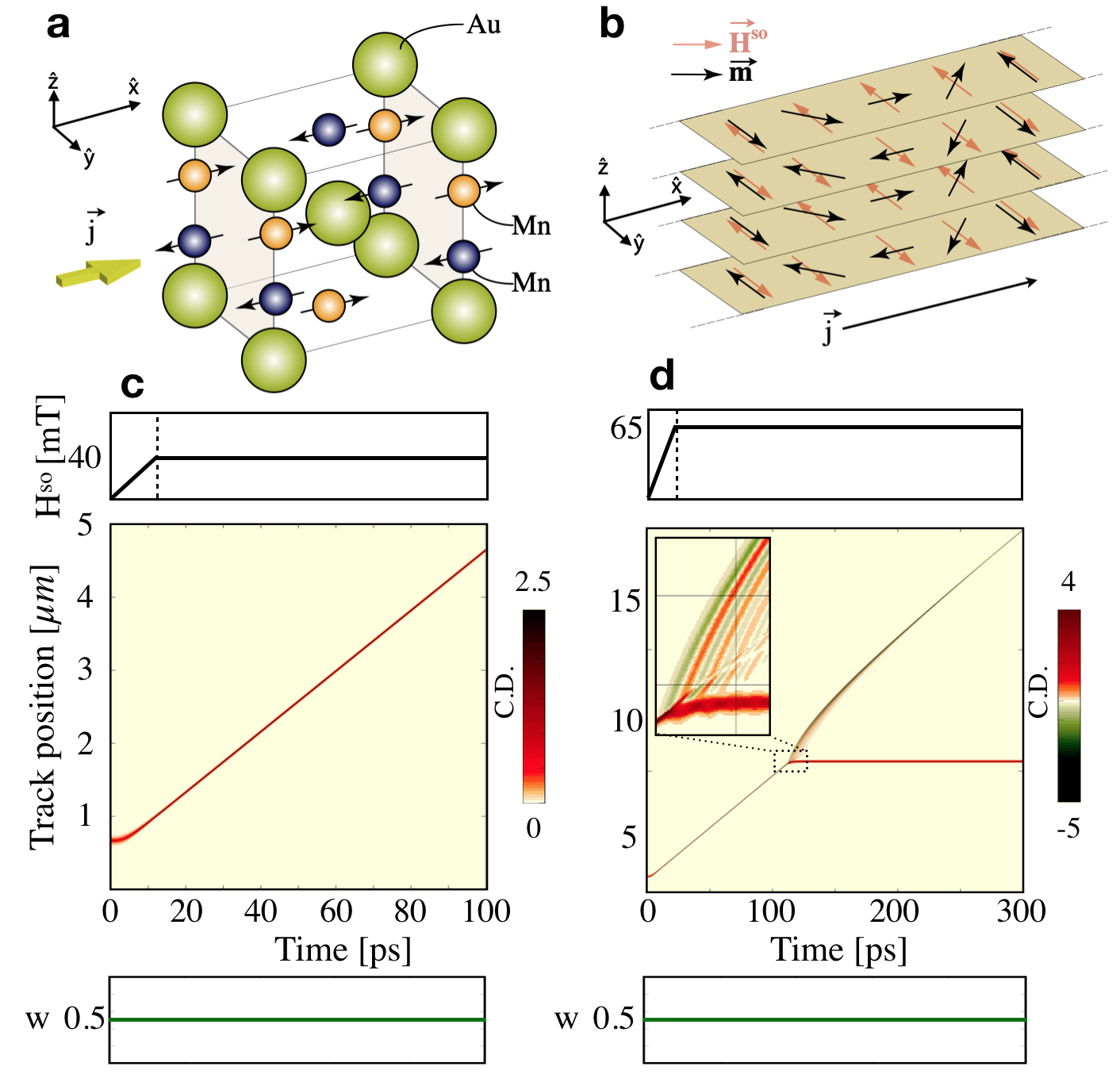}
\caption{\textbf{System of study.} 
\textbf{a} Crystal and spin structure of Mn$_2$Au with a basal plane crystal lattice a$_0$= 0.3228 nm. Yellow arrow represents the current density injection direction ($\hat{x}$-axis) 
\textbf{b} Schematic illustration of each basal plane containing a N\'{e}el type of DW.
\textbf{c} Top panel represents the excitation protocol where the SO-field is ramped up to its maximum value in 10 ps. Middle panel shows the spatiotemporal evolution of the charge density (C.D.) showing that after an initial inertial regime of motion, DW reaches a quasi steady motion. Bottom panel shows the numerically extracted winding number.
\textbf{d} Top panel represents the excitation protocol where the SO-field is ramped up to its maximum value in 10 ps. Middle panel shows the spatiotemporal evolution of the charge density (C.D.) showing the nucleation a DW-pair, appearance of spin-waves and supermagnonic regime of motion of one of the nucleated DWs.  Bottom panel shows the numerically extracted winding number.
  }
\label{fig:fig1}
\end{figure}


\newpage

\begin{figure}
\centering
\includegraphics[scale=0.55]{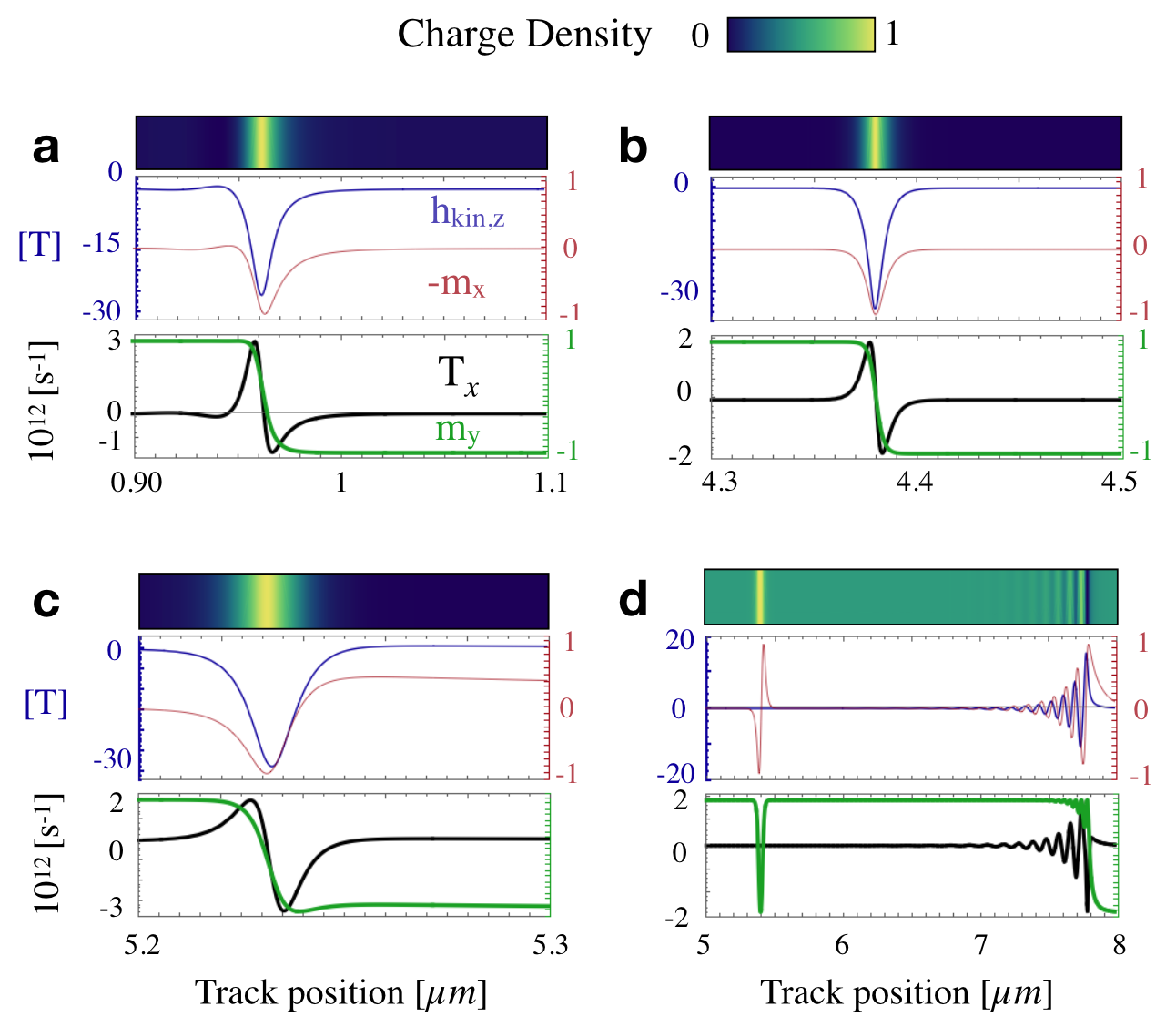}
\caption{
 \textbf{Nucleation process.}
 Snapshots of the magnetisation $x$ and $y$-components, kinetic field $z$-component and $x$-component of the torque exerted by the kinetic field at different instances during the DW$_1$ motion. Panel \textbf{a} to \textbf{d} corresponds to physical times of 10, 90, 110 and 145 picoseconds. At all times during the atomistic spin dynamics simulations the $\hat{x}$ and $\hat{y}$-components of the kinetic field reach values of the order of mT while the $z$-component, $h_{\text{kin,z}}$ can reach absolute values circa 35 T. When the kinetic field is not centred at the DW central position, it induces a DW deformation through the the additional torque it exerts onto the local magnetisation. 
}
\label{fig:fig2}
\end{figure}

\newpage


\begin{figure}
\includegraphics[scale=0.45]{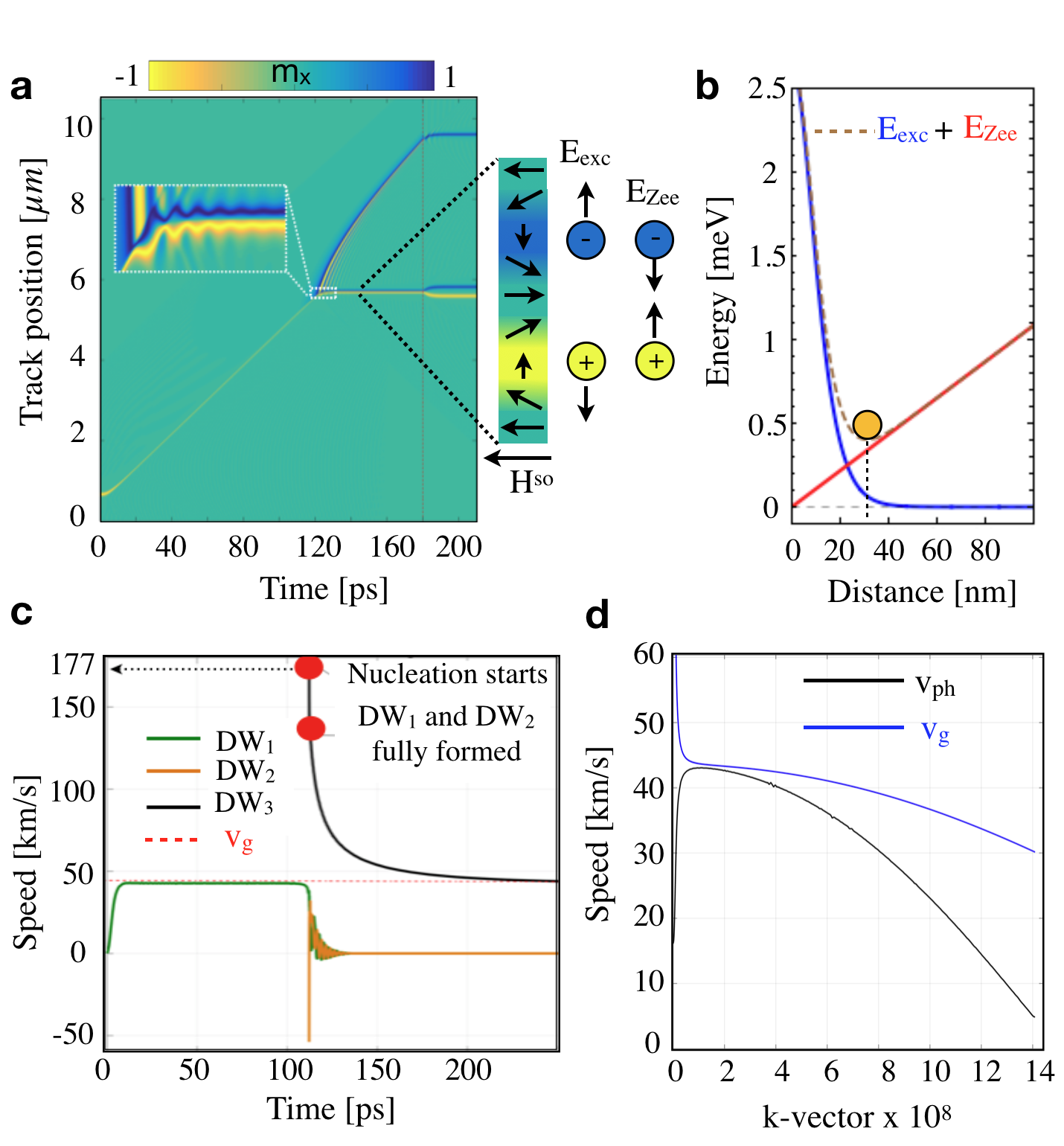}
\centering
\caption{\textbf{DW dynamical properties.} 
\textbf{a} Spatiotemporal evolution of the magnetisation. Inset represents the DW$_1$ and DW$_2$ initial oscillations before reaching a stable distance between them. The separation is due to the competition between the Zeeman energy provided by the SO-field which tries to pull them apart while the exchange tries to separate them as both DWs have the same winding number.  
\textbf{b} Stable distance between DW$_1$ and DW$_2$ of 29 nm calculated for an applied SO-field of 65 mT. 
\textbf{c} DWs velocity as a function of time for each DWs. At the moment the nucleation starts, the maximum registered speed of DW$_3$ travelling at supermagnonic speeds is 177 km/s denoted and 133 km/s when both DWs are fully formed. A smooth recovery of DW$_3$ towards submagnonic speeds is observed and lasts for about 100 ps. 
\textbf{d} Phase and group velocity of the magnons as a function of the wave-vector for an applied SO-field of 65 mT extracted from the dispersion relation. Notice that once DW$_3$ enters into the supermagnonic regime of motion it also surpass the phase velocity of the magnons which is manifested by the spin-Cherenkov radiation.}
\label{fig:fig3}
\end{figure}

\newpage

\begin{figure}
\centering
\includegraphics[scale=0.7]{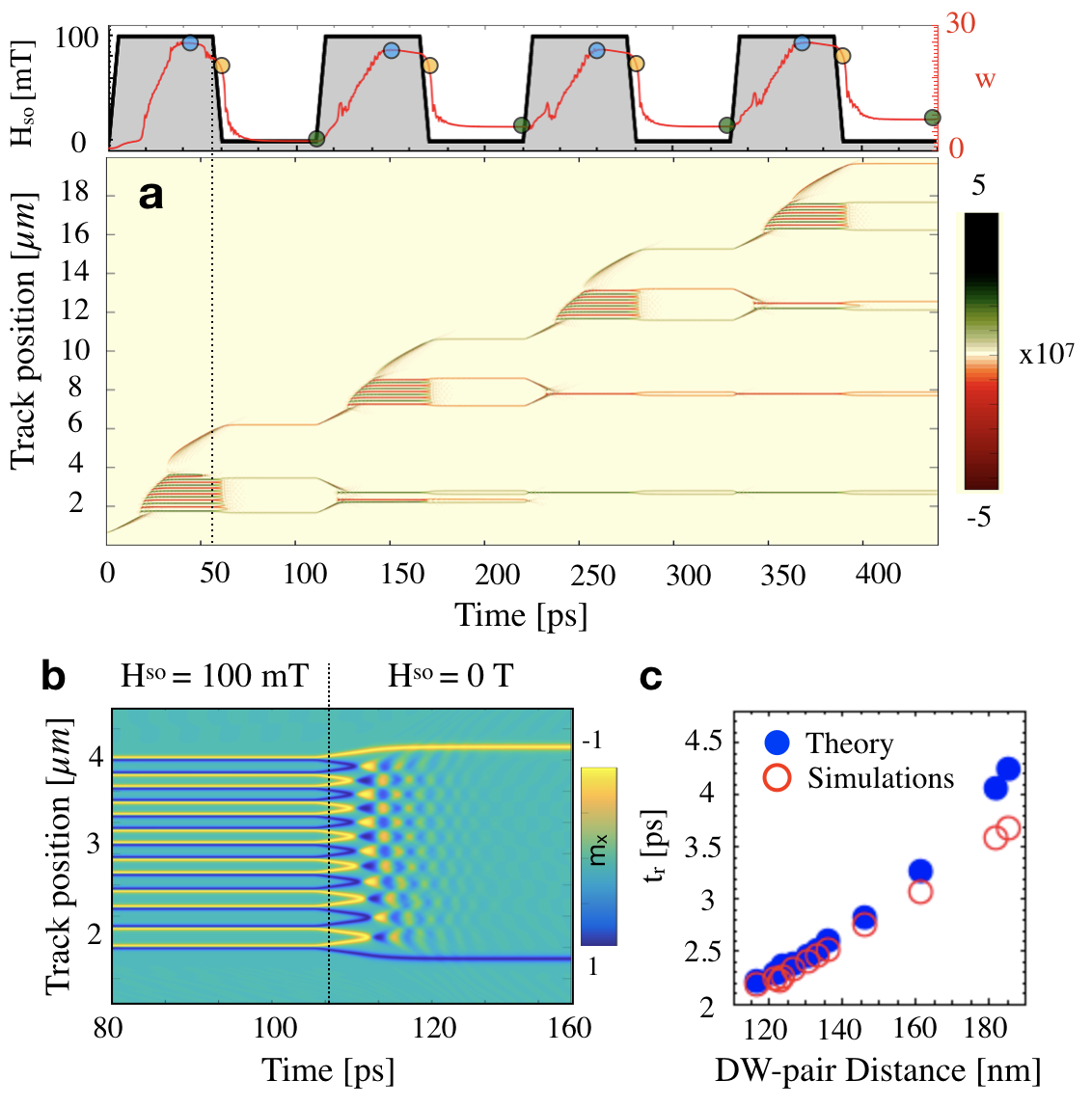}
\caption{\textbf{DW avalanches and recombination.} 
\textbf{a} Top panel show the excitation protocol and the winding number variation versus time. Each of the pulses are characterised by a ramp and fall times of 5 ps and constant SO-field values of 100 mT and 0 T for 50 ps. Below, it is shown the spatiotemporal evolution of the charge density. Multiple DW-avalanches can be observed at each SO-field pulse which translates into an increase in the number of the winding number, w. 
\textbf{b} Time evolution of the $m_x$-component of the magnetisation which illustrates the recombination dynamics between DWs with opposite winding number once the SO-field is set to zero Tesla. 
\textbf{c} Comparison of the recombination time, $t_{\text{r}}$, extracted from the simulation and analytically as a function of the distance among the nearest DWs with opposite winding number shown in panel \textbf{b}.
}
\label{fig:fig4}
\end{figure}

\clearpage

\section*{References}
\bibliography{references}


\end{document}